\documentclass[aps,prfluids,11pt,superscriptaddress,floatfix,tightenlines,showpacs,longbibliography,notitlepage]{revtex4-2}

\usepackage{amsmath,graphicx,amssymb,bm,mathrsfs,amsthm,xparse}
\usepackage{pgfplots}
\usepackage{subcaption}
\usepackage{soul,placeins,lineno}
\pgfplotsset{compat=1.18}
\usepackage{filecontents}
\usepackage{standalone}
\usepackage[dvipsnames]{xcolor}
\usepackage[breaklinks,colorlinks = true,linkcolor = blue,urlcolor=blue,citecolor=blue]{hyperref}

\usepackage{lineno}

\usepackage{comment}

\newcommand{\RbyReqpdf}{R/ R_\mathrm{eq}}
\newcommand{\Nb}{N_\mathrm{b}}
\newcommand{\Nk}{N_\textsc{k}}
\newcommand{\bk}{b_\textsc{k}}

\newcommand{\taus}{\tau_\mathrm{s}}

\newcommand{\h}{h^\star}
\newcommand{\Wi}{\mathrm{Wi}}
\newcommand{\Wistar}{\mathrm{Wi}^\star}
\newcommand{\Wicstar}{\mathrm{Wi}^\star_\mathrm{c}}
\newcommand{\thetabar}{\overline{\theta}}

\newcommand\Wic{\mathrm{Wi}_{\rm c}}
\newcommand\Req{R_{\mathrm{eq}}}
\newcommand\Rmean{\langle R\rangle}
\newcommand\Qeq{Q_{\mathrm{eq}}}
\newcommand\Qm{Q_{\mathrm{m}}}
\newcommand\Rmax{R_{\mathrm{m}}}

\newcommand{\revise}[1]{\textcolor{black}{#1}}

\newcommand{\rp}[1]{\textcolor{black}{#1}}
  
\begin{document}

\title{
How hydrodynamic interactions alter polymer stretching in turbulence}

\author{Aditya Ganesh} 
\affiliation{IITB-Monash Research Academy, Indian Institute of Technology Bombay, Mumbai 400076, India}
\affiliation{Department of Chemical Engineering, Indian Institute of Technology Bombay, Mumbai 400076, India}
\affiliation{Department of Mechanical and Aerospace Engineering, Monash University, Clayton, 3800, Australia}

\author{Dario Vincenzi} 
\affiliation{Universit\'e C\^ote d'Azur, CNRS, LJAD, 06100 Nice, France}

\author{Ranganathan Prabhakar}
\affiliation{Department of Mechanical and Aerospace Engineering, Monash University, Clayton, 3800, Australia}

\author{Jason R. Picardo}
\affiliation{Department of Chemical Engineering, Indian Institute of Technology Bombay, Mumbai 400076, India}

\begin{abstract}
  Hydrodynamic interactions (HI) between segments of a polymer have long been known to strongly affect polymer stretching in laminar viscometric flows. Yet the role of HI in fluctuating turbulent flows remains unclear. Using Brownian dynamics simulations, we examine the stretching dynamics of bead-spring chains with inter-bead HI, as they are transported in a homogeneous isotropic turbulent flow (within the ultra-dilute, one-way coupling regime). 
  \revise{We find that the effects of HI are negligibly small for dumbbells but become increasingly prominent as the polymer model is refined, i.e., as the number of beads are increased.}
 HI-endowed chains exhibit a steeper coil-stretch transition as the elastic relaxation time is increased, i.e., HI cause less stretching of stiff polymers and more stretching of moderately elastic polymers. The probability distribution function of the end-to-end extension is also modified, with HI significantly limiting the range of extensions over which a power-law range appears. 
  On quantifying the repeated stretching and recoiling of chains by computing persistence time distributions, we find that HI delays migration between stretched and coiled states. 
 These effects of HI, which are consistent with chains experiencing an effective conformation-dependent drag, are sensitive to the level of coarse-graining in the bead-spring model. Specifically, an HI-endowed dumbbell, which cannot form a physical coil, is unable to experience the hydrodynamic shielding effect of HI. 
  Our results highlight the importance of incorporating an extension-dependent drag force in dumbbell-based simulations of turbulent polymer solutions. To develop and test such an augmented dumbbell model, we propose the use of a time-correlated Gaussian random flow, in which the turbulent stretching statistics are shown to be well-approximated.

\end{abstract}

\maketitle

\section{Introduction}

A dissolved polymer when stretched exerts elastic feedback forces onto the solvent. At the macroscale, this feedback renders the solution viscoelastic and produces phenomena like turbulent drag reduction \citep{g14,bc18,x19,sbgc22} and inertialess chaotic flow \citep{steinberg21,Dutta22}. At the microscale, that is, at the scale of the polymer, the disturbance flow produced by the feedback force of one part of the polymer can alter the motion of all other parts. Such solvent-mediated interactions between different parts of the polymer are termed hydrodynamic interactions (HI).
Their consequences for polymer stretching and stress have been studied extensively in laminar viscometric and extension-dominated flows \citep{jendrejack02,ssc04,pp04,stsc05,prabhakar-blob-07,pp07,s18}. The effects of HI in turbulent flows, however, are not well understood. With rare exceptions \citep{sg03,kwt10,vwrp21,pv25}, HI have typically been ignored in studies of the Lagrangian dynamics of polymers in turbulence.  \revise{Our goal in this work is to examine the nature and extent of the influence of HI on the stretching of polymers in turbulent fluctuating flows, while considering a hierarchy of bead-spring polymer models.}

\rp{To set the stage, it is helpful to recall the physical consequences of HI. A long flexible polymer may be pictured as a freely-jointed chain of $\Nk$ Kuhn segments, each of length $\bk$. Under quiescent (thermodynamic) equilibrium conditions, the chain forms a random coil with size $R_{\rm eq}$. When such a coil is dragged through a solvent, interior segments are partially \emph{shielded} from the background flow because the disturbance flows created by many segments superpose. In effect, the solvent cannot easily ``drain'' through the coil at the same rate that it would without HI. Hence, the drag experienced by the coil is akin to that of a sphere with an effective hydrodynamic size.
As the polymer stretches out, the geometric configuration changes, shielding weakens, and the effective drag changes accordingly. In extension-dominated laminar flows, this produces an effective conformation-dependent drag \citep{deGennes74,h75,Hinch77,Magda88,Larson97-DNA,sbsc03,ssc04}, which in turn can yield coil-stretch hysteresis \citep{deGennes74,h75,sbsc03,ssc04,Gao2024}: coiled configurations tend to remain coiled longer, and stretched configurations tend to remain stretched longer \citep{ssc04,pp07}. These are not subtle effects in strong extension; they can qualitatively reorganize stretching dynamics by slowing migration between conformational states, and lead to hysteresis in rheological measurements \citep{Prabhakar2006, Sridhar2007, Prabhakar2017-ri}.}

In the context of turbulence one is usually interested in situations {where polymers are highly stretched,}
so that viscoelastic effects are strong enough to produce drag reduction \citep{g14,bc18,x19,sbgc22} in high Reynolds number flows, or generate elasto-inertial \citep{Hof23-AR,Graham2019,Ganesh21,Rosti24} and elastic turbulence \citep{shankarRev22,Dutta22,Morozov23,Kerwell23,rpm21,Garg25} in moderate and low Reynolds number flows. Will HI be relevant in such scenarios, given that the effects of HI are relatively weak for fully-extended polymers? Yes, because in turbulence the distribution of polymer extensions {is always broad, owing to the fluctuating nature of the strain-rate.}

Indeed, the theory of \citet*{bfl00,bfl01} and \citet{c00}, developed {for a dumbbell (without HI)} in a general chaotic flow, 
predicts that the probability distribution function (PDF) of the extension $R$ will always exhibit a power-law behaviour. The corresponding exponent increases with $\Wi$, crossing $-1$ at $\Wi = 1/2$, when $\Wi$ is defined as the product of the polymer relaxation time and the Lyapunov exponent of Lagrangian trajectories in the flow.
Since a dependence of $R^{-1}$ implies a non-normalizable PDF, in the idealized case of a Hookean dumbbell, $\Wi = 1/2$ is treated as the critical point of a coil-stretch transition. This is not mere convention---the PDF of $R$ does indeed exhibit hallmark features of phase transitions near $\Wi = 1/2$, such as a slowing down of dynamics \citep{mav05,cpv06,ppv23} and a maximization of entropy \citep{ssflls21,Vincenzi-entropy23}. The broad, power-law PDF of $R$, which has been measured in experiments \citep{gcs05,ls10,ls14} and clearly observed in simulations of free-draining (no HI) dumbbells and chains
\citep{wg10,bmpb12,rll22,sbgc22,ppv23}, implies that many polymers will be coiled even at large $\Wi$. When a stretched polymer is advected into a region of the flow with low straining, it will recoil and remain coiled until it encounters high strain-rates (the residence time in coiled states follows a Poisson distribution \citep{ppv23}). This is when HI will be important. Indeed, we show here that HI delay the migration between coiled and stretched states, and thereby alter the PDF of extension qualitatively. 

In this work, we study the effects of HI on polymer stretching in homogeneous isotropic turbulence, using a bead-spring chain model for the polymer. \rp{Ideally, one would like to simulate the polymer at Kuhn-segment resolution using a bead-rod chain; however, such simulations with HI in turbulence are impractical at present, especially when large ensembles of long trajectories are required. We therefore resort to the coarse-grained bead-spring description in which the bead-rod chain is divided into $\Nb$ subchains, each represented by an isotropic ``bead'' connected by entropic springs. }

\rp{In the Rouse model, each bead has an isotropic Stokes friction coefficient $\zeta$, and the HI between different beads are neglected. As a result, the overall friction of a Rouse chain is simply $\Nb \,\zeta$ and the drag on the chain is independent of the chain's conformation.} 
\rp{When one retains inter-bead HI, for example via the Rotne--Prager--Yamakawa (RPY) mobility tensor, then the model acquires the ability to represent shielding at the scale of bead-bead separations. 
The chain will therefore experience a conformation-dependent drag, owing to shielding at equilibrium and deshielding as the chain uncoils and stretches out.  Importantly, the extent to which a coarse-grained model can represent this conformation dependence varies with $\Nb$: a few widely spaced beads may under-resolve shielding effects that a finer representation (many closely spaced beads) can capture. This connection between the effects of HI and coarse-graining is well established for simple rheometric flows~\citep{diaz89,pps04,prabhakar-blob-07}, wherein it is clear that a dumbbell with HI fails to capture the effects of HI on a more realistic, multi-bead chain. Therefore, to determine the influence of HI on polymer stretching in turbulence, it is not sufficient to merely add HI to a dumbbell model. A weak effect of HI on dumbbell stretching (a result obtained in recent analytical work~\citep{pv25} and supported by our simulations) need not imply that HI is unimportant but only that a more refined polymer model (with more beads) may be needed to reveal the full effects of HI. Our work shows that the latter is true.}
The bead-spring chain used in our study is described in Sec.~\ref{sec:model}. For simplicity, we ignore excluded-volume interactions and focus on HI alone. \rp{In this section, we describe how the parameters of the chain are varied with $\Nb$ in order for chains with increasing $\Nb$ to correspond to increasingly refined descriptions of the same physical polymer.} We also discuss the turbulent carrier flow here. In Sec.~\ref{sec:db-vs-chain}, we show that HI effects are qualitatively different for dumbbells and multibead chains: a dumbbell cannot bend into a physical coil and therefore does not experience shielding in the same manner as a chain; \rp{accordingly, weak sensitivity to HI at $\Nb=2$ does not imply a weak sensitivity at higher $\Nb$.} In Sec.~\ref{sec:stats} we show that HI steepens the coil-stretch transition for chains and qualitatively modifies the extension PDF at small and intermediate $R$, reducing the range over which a clear power-law is visible. By analyzing the persistence time of chains in stretched and coiled configurations, we directly confirm that these effects of HI are a consequence of HI-endowed chains taking longer to migrate between states of small and large extensions. Finally, we check whether the effects of HI on turbulent stretching can be predicted using a random model for the turbulent velocity gradient. Our results, in Sec.~\ref{sec:random}, show that a time-correlated Gaussian random model works well, suggesting its use as a testing ground for developing coarse-grained polymer models, capable of capturing the effects of HI in fluctuating flows. \revise{We end in Sec.~\ref{sec:conclusion} where we discuss the implications of our results for the simulation of viscoelastic turbulence and suggest directions for future work.}


\section{Brownian chains and fluctuating velocity gradients}\label{sec:model}

\subsection{Bead-spring chain with hydrodynamic interactions}\label{sec:chain}

We consider a freely-jointed chain of $\Nb$ beads, whose evolution is described in terms of its center of mass, $\bm X_c$, and the separation
vectors between beads, $\bm Q_i$ ($i=1,\dots,\Nb-1$) \citep{bird,o96,ssc04}:%
{
\begin{align}
  \label{eq:cm}
  {\rm d}{\bm X}_c &= \bm u(\bm X_c(t),t)\; {\rm d}t+\dfrac{1}{\Nb}\sqrt{\frac{\Qeq^2}{6\taus}}
  \sum_{i=1}^{\Nb}{\rm d}\bm W_i(t), 
  \end{align}%
  \begin{align}
{\rm d}{\bm Q_i} &=
  \bm\kappa(t)\cdot\bm Q_i \; {\rm d}t +\dfrac{1}{4\taus}
  		\sum_{j=1}^{\Nb} \left(\bm D_{i+1,j}- \bm D_{i,j} \right) \cdot 
        \bm F_{j}^{\rm E}  \; {\rm d}t
      +\sqrt{\dfrac{\Qeq^2}{6\taus}}
  		\sum_{j=1}^{i+1} \left(\bm B_{i+1,j}- \bm B_{i,j} \right) \cdot {\rm d}\bm W_{j}(t).  \label{eq:q}
\end{align}%
}%
A simple measure of the deformation of the chain is provided by the end-to-end separation or extension vector $\bm R=\sum_{i=1}^{\Nb-1}\bm Q_i$; the magnitude $\vert\bm R\vert$ is called the end-to-end extension and is denoted by $R$.

The links between neighboring beads are phantom FENE springs with spring constant $H$. If $\zeta$ is the Stokes drag coefficient of the beads, then each link is characterized by an elastic time scale
$\taus=\zeta/4H$. The net spring force exerted on bead $i$ is $\bm F_{i}^{\rm E}=f_i\bm Q_i-f_{i-1}\bm Q_{i-1}$, where the FENE interaction coefficients 
$f_i=\left({1-\vert\bm Q_i\vert^2/\Qm^2}\right)^{-1}$
ensure that the extension of each link does not exceed its maximum length $\Qm$. The contour length or maximum length of the chain is then given by $\Rmax = (\Nb-1)\Qm$.  {[Note that in Eq.~\eqref{eq:q}, one must set $\bm Q_0=\bm Q_{\Nb}=0$ in the equations for ${ {\rm d}\bm Q_1}$ and ${\rm d} {\bm Q}_{\Nb-1}$.]}

The Brownian forces that act on the beads are
represented by {independent increments of vectorial Wiener processes
${\rm d}\bm W_i(t)$}.
The competition between Brownian and elastic forces sets the equilibrium length scale $\Qeq=\sqrt{3k_BT/H}$, where $k_B$ is the Boltzmann constant
and $T$ is the temperature. $\Qeq$ is the root-mean-square (r.m.s.)
extension of the link at equilibrium in a still fluid, in the Hookean limit of $\Qeq \ll \Qm$. The corresponding r.m.s.~end-to-end extension of the chain at equilibrium is 
$\Req = \Qeq\sqrt{\Nb-1}$ \citep{bird}. 

The drag exerted on the beads by the turbulent velocity field $\bm u$ advects and stretches the chain. The extent of stretching depends on the differences in fluid velocity sampled by neighbouring beads, which in turn is written in terms of the velocity gradient at the centre of mass, $\kappa_{ij}=\nabla_j u_i$. This first-order Taylor series expansion of the velocity difference is permissible provided the contour length of the chain, $\Rmax = ({\Nb-1}) \Qm$, is smaller than the viscous Kolmogorov length scale of the flow (as is typically the case \citep{idakcs02,sg03,za03,gsk04,tdmsl04,ps07}).  

Inter-bead hydrodynamic interactions are taken into account via the Rotne-Prager-Yamakawa mobility tensor $\bm D_{i,j}$, defined as 
\begin{align}
\bm D_{i,j} &= \bm I \quad {\rm if} \; i=j,\\
\bm D_{i,j} &= \frac{6a}{8 X_{ij}} \left[ \left( 1+\frac{2a^2}{3X_{ij}^2}\right)\bm I +  
\left( 1-\frac{2a^2}{X_{ij}^2}\right) \frac{\bm X_{ij} \bm X_{ij}}{X_{ij}^2} \right] \quad {\rm if} 
\; i\ne j \; {\rm and} \; X_{ij} \geq 2a,\\
\bm D_{i,j} &= \left[ \left( 1-\frac{9 X_{ij}}{32 a}\right)\bm I +  
 \frac{3}{32} \frac{ \bm X_{ij} \bm X_{ij}}{a X_{ij}} \right] \quad {\rm if} 
\; i\ne j \; {\rm and} \; X_{ij} < 2a,
\end{align}
where $\bm X_i$ is the position vector of bead $i$, $\bm X_{ij} = \bm X_j - \bm X_i$ is the displacement vector between beads $i$ and $j$, and $X_{ij}=|\bm X_{ij}|$. Further, $\bm I$ is a $3 \times 3$ identity tensor, and $a$ is the radius of the beads which defines the non-dimensional hydrodynamic interaction parameter $h^\star$:
\begin{equation}\label{eq:h}
\h=\frac{a}{\Qeq}\left(\frac{3}{\pi}\right)^{1/2}
\end{equation}
To be physically meaningful, $a$ should be less than $\Qeq/2$, else the beads will overlap at equilibrium; therefore, $\h$ should be $\lessapprox 0.49$ \citep{ssc04}. 

Knowing the positive definite mobility tensor $\bm D_{i,j}$, one may compute the coefficient matrix $\bm B_{i,j}$ of the Brownian forces in Eq.~\eqref{eq:q} as follows: 
\begin{equation}
\bm D_{i,j} = \sum_{l=1}^{\Nb} \bm B_{i,l} \cdot \bm B_{j,l}^{\rm T},
\label{eqn:Bij}
\end{equation}
where the superscript $\rm T$ denotes the transpose. Following \citet{jendrejack02}, we compute $\bm B_{i,j}$ by first combining the $\Nb^2$ different $\bm D_{i,j}$ matrices into a $3\Nb \times 3\Nb $ block matrix, and then performing a Cholesky decomposition to obtain a lower triangular block matrix that yields the $\bm B_{i,j}$ matrices. Note that $\bm B_{i,j} = 0$ if $j > i$.

Equations \eqref{eq:cm} to \eqref{eqn:Bij} constitute the bead-spring chain model with hydrodynamic 
interactions. For a given number of beads, $\Nb$, the chain is parameterized by $\Qeq$, $\Qm$, $\taus$, and the HI parameter $\h$ (setting $\h = 0$ yields the free-draining Rouse model \citep{bird,o96}).

\subsection{Dependence of chain parameters on the number of beads}\label{sec:mapping}

\revise{In this work, we consider a polymer with a fixed contour length or maximum extension $\Rmax$ and model it as a bead-spring chain, so that the number of beads represents the level of coarse-graining.
To understand how the effect of HI depends on coarse-graining,} 
we need a way to compare the results of chains with different numbers of beads. In the absence of HI, the mapping of \citet{jc07} has been found to work very well in homogeneous isotropic turbulent flows. Specifically, the PDF of $R/\Req$ of two chains with different values of $\Nb$ are found to be in agreement, provided the parameters are determined in the following manner. First, we fix the equilibrium length of the links $\Qeq$. {This choice also fixes the equilibrium length of the dumbbell version of the model, $\Req^D = \Qeq$.} Next, we choose values for the maximum extension and the elastic time scale for the dumbbell model: $\Rmax^D$ and $\tau^D$. We then calculate the remaining link parameters, $\Qm$ and $\taus$, for a chain with $\Nb$ beads according to 
{
\begin{equation}\label{eq:jcmap}
   \frac{\Qm}{\Qeq} = \frac{1}{\sqrt{\Nb-1}}\,\frac{\Rmax^D}{\Req^D},\quad \taus = \frac{\quad 6 \tau^D}{(\Nb+1)\Nb}.
\end{equation}
}%
This mapping is consistent with the view that the chain is a coarse-grained representation of a polymer with a fixed number of Kuhn steps $\Nk$. Recalling that ${\Nk}^{1/2}$ is proportional to the ratio of the contour length of the chain $\Rmax$ to its equilibrium r.m.s.~extension $\Req$, we see that maintaining $\Nk$ constant as we coarse-grain requires $\Rmax/\Req$ to be constant. Now, for a fixed $\Qeq$, we have $\Req = \sqrt{\Nb-1} \Qeq$ (based on the random-walk theory for a Rouse chain \citep{deGennes79,bird}). So, the mapping \eqref{eq:jcmap} yields $\Rmax/\Req = (\Nb-1) \Qm/\sqrt{\Nb-1} \Qeq = \Rmax^D/\Qeq$, which is independent of $\Nb$ as we have fixed the values of both $\Rmax^D$ and $\Qeq$. Thus, the Jin-Collins mapping sets the link parameters so that the chain's contour length $\Rmax$ increases as $\sqrt{\Nb-1}$, thereby ensuring that $\Rmax/\Req$ is independent of $\Nb$. Since $\Rmax/\Req = \Rmax^D/\Qeq$, we see that the initial choices for $\Rmax^D$ and $\Qeq$ sets the extensibility of the polymer for all levels of coarse graining, including that of the dumbbell (for which $\Req^D = \Qeq)$. In our simulations, we set $\Qeq=1$ and $\Rmax^D=109.54$, yielding an extensibility ratio of ${\Nk}=\Rmax^2/\Req^2 = 12000$. While being relevant to long DNA molecules \citep{sbsc03}, {which have been used in drag-reduction experiments \citep{Choi02,Lim2003},} this value of the extensibility ratio ensures a sufficient range of extension for a power-law distribution to manifest.

The time-scale $\tau^D$ in \eqref{eq:jcmap} may be thought of as an approximation to the longest relaxation time of the entire chain. Its relation to the time-scale of the individual springs $\taus$ is consistent with the behavior of the largest relaxation time of a Rouse chain {(no HI)}, which for large $\Nb$ scales as $\Nb^2 \taus$. The precise form of the relation between $\tau^D$ and $\tau^s$ was obtained by \citet{jc07} on substituting the relation between $\Req$ and $\Qeq$ into an expression for the elongational viscosity of highly-stretched polymers, given by \citet{wt89}, and requiring the resulting viscosity values to be independent of $\Nb$. 

{In the absence of HI, the Jin-Collins mapping has been found to work very well for bead-spring chains in homogeneous isotropic turbulent flows: on using the mapping, the PDF of $R/\Req$ of a dumbbell matches that of a chain with 10 or 20 beads.} This is true, not only for the stationary PDF \citep{wg10}, but also for the evolving PDF that describes the initial stretching of polymers when they are first introduced into a turbulent flow \citep{ppv23}. {Here, we use the Jin-Collins mapping as a basis of comparison that is particularly well-suited to revealing the effects of HI, given that it accounts for changes due to varying $\Nb$ in the absence of HI.}

{The value of $h^\star$ is, in principle, tied to the value of $\taus$ through the bead radius $a$ [see \eqref{eq:h} and recall that $\taus =\zeta/4H$ where $\zeta = 6\pi \mu a$, where $\mu$ is the solvent viscosity]. } In practice, however, this relationship is often ignored and $h^\star$ is treated as a free parameter that controls the strength of HI \citep{Hsieh03,prabhakar-blob-07}; typically, its value is tuned to reproduce experimental observations, e.g., of the longest relaxation time \citep{Hsieh03}. Here, {we vary $h^\star$ freely
in order to understand the influence of HI on the stretching dynamics of the chain}; specifically, we use 
$h^\star = 0$ (free-draining), $h^\star = 0.2$ (an intermediate and typical value \citep{ssc04,pp04}), and $h^\star = 0.49$ (the largest physically meaningful value).



\subsection{Brownian dynamics simulations}\label{sec:BD}

 From Eq. \eqref{eq:cm}, we see that the centre of mass of the chain moves like a tracer in the turbulent flow, since its Brownian motion is much weaker than turbulent advection. Further, as we are focusing on the
the ultra-dilute limit in which the flow remains unaffected by elastic feedback forces, the tracer-trajectory of a chain's centre of mass is independent of its internal structure. Therefore, Eq.~\eqref{eq:q} for the chains extension may be solved independently, given the Lagrangian data of the velocity gradient $\bm\kappa(t)$ along a tracer trajectory. Here, we use pre-stored Lagrangian data from a direct numerical simulation (DNS) of homogeneous isotropic incompressible turbulence, as well as a random Gaussian model for $\bm\kappa(t)$ (see Sec.~\ref{sec:turbflow}).

Equation~\eqref{eq:q} is integrated using the Euler-Maruyama method. 
The rejection algorithm of Ref.~\cite{o96} is used to prevent $Q$ from exceeding $\Qm$ and causing a numerical divergence of the FENE force. 
Any time-step update that yields extensions greater than \revise{ $\Qm\big(1-\sqrt{\delta t/\taus}\big)^{1/2}$ is rejected \cite{o96}.} By using sufficiently small time steps $\delta t$, we have limited such rejections to a negligible fraction of the total steps in a simulation. Note that, unlike the case of extensional flow, polymers in turbulence do not persistently align with the straining direction of the velocity gradient due to constant vorticity-induced rotation. Hence, sophisticated implicit time-integration methods are not needed here. As a check, though, we have redone a few simulations, with and without HI, using the semi-implicit, predictor-corrector adaptive time-stepping algorithm of \citet{pp04}, and find that our results for the PDF of extension remain unchanged.

We perform simulations of chains with $\Nb$ equal to 2 (dumbbells), 4, 10, and 20. The computational cost of performing Brownian dynamics simulations, over a large ensemble of trajectories in the turbulent flow, limits us to twenty-bead chains.

We consider a range of values of the Weissenberg number, defined as $\Wi = \lambda \tau^D$.
The Lyapunov exponent $\lambda$, which is computed along tracer trajectories in the turbulent carrier flow (Sec.~\ref{sec:turbflow}),
characterizes the long-time asymptotic exponential-stretching behaviour of line elements. Therefore $\lambda^{-1}$ is the most suitable flow time-scale for defining $\Wi$. Indeed, with this definition, the distribution of polymer extensions undergoes a coil-stretch transition near $\Wi \approx 1/2$, as was first anticipated theoretically by \citet{bfl00,bfl01} and \citet{c00}, and later demonstrated for free-draining dumbbells and chains in homogeneous isotropic turbulence \citep{Vincenzi-entropy23,ppv23}.

\revise{For every combination of parameter values we perform Brownian dynamics simulations of $2\times10^5$ independent trajectories, each of temporal length $75 \lambda^{-1}$ (longer computations up to $300 \lambda^{-1}$ are performed for the persistence-time analysis of Sec.~\ref{sec:persistence}). }
Each chain starts out with a random configuration and is allowed to equilibrate in a still fluid ($\bm \kappa = 0$). The flow is then 'turned on', with $\bm \kappa$ drawing values from either the DNS dataset or the random model. Since the distribution of chains starts out from an equilbrium distribution, some time is required for the chains to stretch out in the fluctuating flow and for the stationary distribution of extension to be established. In this work, we ignore the transients and focus on the stationary statistics.
\revise{The time step of the simulations is varied according to $\Wi$ so that $\delta t \approx 0.05 \taus$ (e.g., for $\Wi = 1$, $\delta t = 7.5 \times 10^{-4} \lambda^{-1} =5.5 \times 10^{-3} \tau_\eta = 0.05 \taus$). }

\subsection{Velocity gradients in turbulent and random flows}\label{sec:turbflow}

To study polymer stretching in homogeneous isotropic turbulence, we use a database of velocity gradients $\bm\kappa(t)$, stored along the trajectories of $2 \times 10^5$ tracers. The corresponding DNS, with Taylor-microscale Reynolds number $\mathit{Re}_\lambda\approx 111$, was performed in a tri-periodic cube, wherein the incompressible Navier--Stokes equations were solved using a standard fully-dealiased pseudo-spectral method with $512^3$ grid points \citep{jr17}. Time integration was performed using a second-order Adams--Bashforth scheme. \revise{The flow was driven to a statistically stationary
state via time-dependent large-scale forcing that injected energy at a fixed rate into the first and second wavenumber shells. The corresponding Fourier modes of the instantaneous velocity field were summed and then multiplied by a time-dependent factor to yield the forcing function; the instantaneous value of the factor is determined by the ratio of the desired energy injection rate and the instantaneous kinetic energy content of the first two wavenumber shells.} The tracers were evolved in the stationary flow using a second-order Runge--Kutta method \citep{jr17}. We have calculated the Lyapunov exponent $\lambda$ along these tracer trajectories from $\bm \kappa(t)$, using the continuous $QR$ method \citep{gpl90,drv97,mcdh01}; in terms of the Kolmogorov time-scale, we obtain $\lambda=0.136/\tau_\eta$ in keeping with previous simulations of isotropic turbulence \citep{bbbcmt06}.  Further details on the DNS data set are provided in Appendix~\ref{sec:turb-details}.

We also check whether polymer stretching can be predicted using a Gaussian random model for $\bm \kappa(t)$. While non-Gaussian, extreme straining events will be lost, other important statistical properties of the Lagrangian velocity gradient can be imitated by a Gaussian model. Specifically, we use the random model of \citet*{bkl97} to generate a time series for $\bm \kappa(t)$ that has the same Lyapunov exponent $\lambda$ as the turbulent velocity gradient, as well as approximately the same temporal correlations of Lagrangian vorticity and strain-rate. The model and its tuning is described in Appendix~\ref{sec:turb-details}. 
This correlated Gaussian model was shown by \citet{ppv23} to be a good surrogate for the turbulent velocity gradient in terms of reproducing the stretching statistics of Rouse chains. We will see in Sec.~\ref{sec:random} that the random model also reproduces the effects of HI on turbulent polymer stretching.  
For a given polymer and solvent, the value of $\Wi$ can be increased by intensifying the turbulence, so that $\lambda$ increases. Consider a stationary homogeneous isotropic turbulent flow that is driven by a large-scale forcing, which injects energy at a rate $\epsilon$. The turbulent properties of the flow are entirely characterized by the Reynolds number $Re = UL\rho/\mu$, or alternatively by the Taylor Reynolds number $Re_\lambda \sim Re^{1/2}$. Here, $U$ is the r.m.s. velocity, $L$ is the size of the largest eddies, and $\rho$ is the density of the fluid. Now, $\lambda$ is proportional to the inverse of the Kolmogorov time-scale $\tau_\eta$ \citep{bbbcmt06}, which is the turnover time of the smallest eddies in the flow. Using the definition $\tau_\eta = (\mu/\rho \epsilon)^{1/2}$, and the \revise{Taylor dissipation law} $\epsilon \sim U^3/L$, one 
has $\tau_\eta = (\nu L)^{1/2}U^{-3/2} \sim (L/U) Re_\lambda^{-1}$ \citep{f95}. Usually, $L$ is set by the geometry of the system or the forcing mechanism, so that as one increases the forcing strength, $U$ will increase while $L$ will not. As a result, $\tau_\eta$ will decrease and $\lambda$ and $\Wi$ will increase; in addition, $Re_\lambda$ will increase. 
\revise{Our simulations, however, are performed at a single fixed value of $Re_\lambda$. So, strictly speaking, our simulations correspond to an experimental scenario in which $\Wi$ is increased, while keeping $Re_\lambda$ constant,} by adjusting the forcing so that $U$ increases while $L$ decreases (leaving $UL$ unchanged).
Nonetheless, we expect our results to hold to a good approximation even if $Re_\lambda$ increases along with $\Wi$. As $Re_\lambda$ increases, the distribution of strain-rates will become more heavy-tailed, i.e., the polymer will experience more extreme straining events \citep{Buaria2019,Buaria2022}. However, non-Gaussian extreme strain-rates have only a weak effect on polymer stretching; this result was shown for Rouse chains by \citep{ppv23} and is verified for chains with HI in Sec.~\ref{sec:random}. So, our results on the effect of increasing $\Wi$ will be relevant even to situations in which the increase of $\Wi$ is accompanied by an increase in $Re_\lambda$. 

\revise{As noted in Sec.~\ref{sec:chain}, our Brownian dynamics simulations assume a linear velocity field in the vicinity of the polymer which in turn requires $\Rmax < \eta$ (where $\eta = (\nu^3/\epsilon)^{1/4}$ is the Kolmogorov length scale). This condition sets limits on the range of physical systems to which our results can be applied. For e.g., for a particular polymer and solvent combination, with given values of $\Rmax$ and $\nu$, the requirement that $\Rmax < \eta$ sets an upper limit on the range of dissipation rates $\epsilon$; of course, by varying the solvent viscosity one can alter the permissible range of $\epsilon$. Alternatively, if the solvent and dissipation rates are fixed, then the condition $\Rmax < \eta$ limits the size of polymers to which our results may be applied (effectively limiting $\Req$ since our simulations use $\Rmax^2/\Req^2 = 12000$).}

\section{Dumbbells and chains respond differently to HI}\label{sec:db-vs-chain}

We begin by examining how HI affects the mean extension of chains with different numbers of beads. Fig.~\ref{fig:avgR-Wi}(a) presents $\Rmean/\Req$ for various values of $\Wi$ for dumbbells ($\Nb = 2$), where the average is calculated over time (in the stationary state) and over all trajectories in the turbulent flow. The same results for chains with $\Nb = 20$ are presented in Fig.~\ref{fig:avgR-Wi}(b). Clearly, both dumbbells and chains undergo a coil-stretch transition as $\Wi$ is increased. For the moment, we do not adjust the definition of the Weissenberg number to account for the change in the relaxation time due to the inclusion of HI. So, rather than examining the variation of extension with $\Wi$, let us first consider how the qualitative effect of HI changes with the number of beads.

\begin{nolinenumbers}
\begin{figure}
\centering
\begin{subfigure}[t]{0.38\textwidth}
    \subcaption{}
    \centering
    \includegraphics{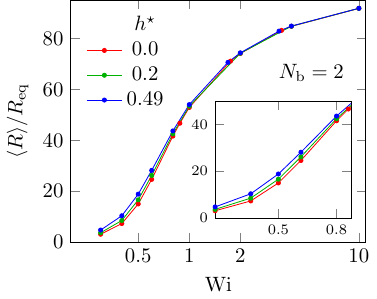}
  \end{subfigure}
  \hspace{.5cm}
    \begin{subfigure}[t]{0.38\textwidth}
    \subcaption{}
    \centering
    \includegraphics{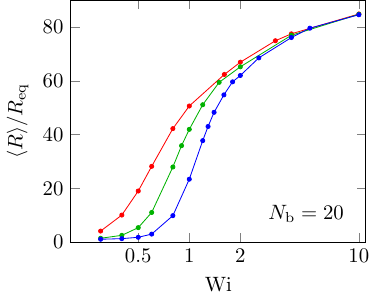}
  \end{subfigure}
  \hfill
  \begin{subfigure}[t]{0.38\textwidth}
    \subcaption{}
    \centering
    \includegraphics{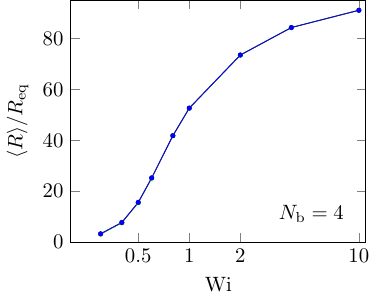}
  \end{subfigure}
  \hspace{.5cm}
  \begin{subfigure}[t]{0.38\textwidth}
    \subcaption{}
    \centering
    \includegraphics{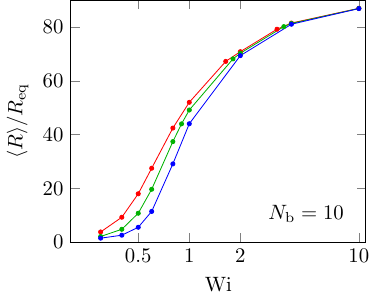}
  \end{subfigure}
\caption{Mean rescaled extension $\langle R\rangle/R_{eq}$ for various values of the Weissenberg number $\Wi$, considering $\h$ = 0, 0.2, 0.49 [see the legend in panel (a)]. Results are presented for (a) dumbbells, (b) twenty-bead chains. (c) four-bead chains, and (d) ten-bead chains. The inset in panel (a) is a zoom of the main panel. The three curves in panel (c) almost entirely overlap.}
\label{fig:avgR-Wi} 
\end{figure}
\end{nolinenumbers}

On comparing the results without HI ($\h = 0$) to those with HI ($\h = 0.2$ and $0.49$), we see that HI marginally \textit{increases} the stretching of dumbbells [Fig.~\ref{fig:avgR-Wi}(a)], while it strongly \textit{decreases} the stretching of twenty-bead chains [Fig.~\ref{fig:avgR-Wi}(b)]. The effect of HI changes gradually from weak promotion to strong suppression of extension, as $\Nb$ is increased from 2 to 20, as evidenced by the results for $\Nb$ = 4 and 10 in Fig.~\ref{fig:avgR-Wi}(c) and Fig.~\ref{fig:avgR-Wi}(d), respectively. The slight enhancement of extension in case of dumbbells is in agreement with recent analytical results, derived using the Batchelor-Kraichnan model of turbulent transport~\citep{pv25}. 

\begin{nolinenumbers}
\begin{figure}
\centering
\begin{subfigure}[t]{0.37\textwidth}
    \subcaption{}
    \centering
    \includegraphics{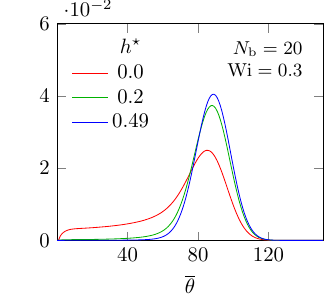}
\end{subfigure}
\hspace{0.5cm}
\begin{subfigure}[t]{0.37\textwidth}
    \subcaption{}
    \centering
    \includegraphics{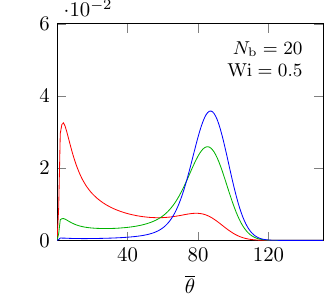}
\end{subfigure}\\
\begin{subfigure}[t]{0.37\textwidth}
    \subcaption{}
    \centering
    \includegraphics{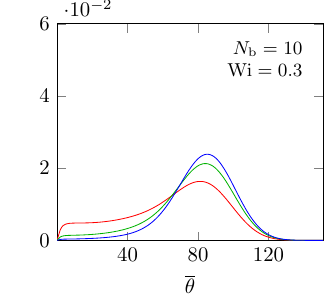}
\end{subfigure}
\hspace{0.5cm}
\begin{subfigure}[t]{0.37\textwidth}
    \subcaption{}
    \centering
    \includegraphics{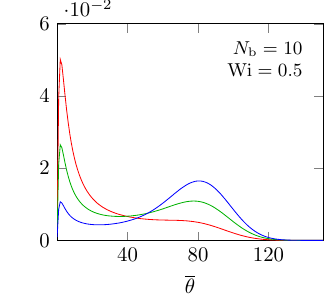}
\end{subfigure}

\caption{PDF of the mean interlink angle $\thetabar$, measured in degrees, for (a,b) twenty-bead chains and (c,d) ten-bead chains. For each case, results are shown for $\Wi = 0.3$ and 0.5, and for $\h = 0, 0.2, 0.49$ [see the legend in panel (a)].}
\label{fig:pdf-theta}
\end{figure}
\end{nolinenumbers}

So why do HI affect dumbbells and multi-bead chains differently? \revise{A plausible explanation is provided by the mechanism of hydrodynamic shielding.} At small $\Wi$, chains retract into coiled configurations in which the outer beads hydrodynamically shield the inner beads and thereby weaken the effect of fluid drag on the chain. As a consequence, small $\Wi$ chains stretch less in the presence of HI. Now, a dumbbell cannot form a coil and is by definition always straight. In this constrained arrangement, the effect of the disturbance flow created by the two beads is to increase the apparent relaxation time scale (as is analytically demonstrable \citep{pv25} via the consistent-preaveraging approximation \citep{o85}). To check whether the influence of HI on multibead chains is associated with the formation of physical coils, we measure the hinge angles $\theta_i$ ($i = 1,2,\dots, \Nb-2$) between successive links, and then compute the average angle $\overline{\theta} = \left(\sum_1^{\Nb-2} \theta_i\right)/(\Nb-2)$ for each chain. \revise{Note that a small end-to-end extension for a multibead chain could result from the links shrinking while the chain remains straight, or from the links retaining their length while the angles vary so that the chain folds into a physical coil. The difference between these two configurations will appear in the value of $\thetabar$, which will be, respectively, near to and far from $0^\circ$ for straight and coiled chains.} The PDF of $\overline{\theta}$ is presented for twenty-bead chains, for $\Wi = 0.3$ and $0.5$, in Figs.~\ref{fig:pdf-theta}(a-b). Increasing $\h$ increases the proportion of configurations with average inter-link angles close to $90^\circ$, which is consistent with HI acting to maintain chains in coiled configurations. 

The shift in the PDF of $\overline{\theta}$ towards $90^\circ$ with increasing $\h$ is not as pronounced when the number of beads is reduced, as demonstrated by the results for $\Nb = 10$ in Fig.~\ref{fig:pdf-theta}(c-d). So, with fewer beads, the shielding effect of HI is weaker; this explains why the reduction in stretching caused by HI is significantly less for ten-bead chains when compared to twenty-bead chains [Figs.~\ref{fig:avgR-Wi}(b,d)]. For four-bead chains, the shielding effect is very weak and just sufficient to counter the opposing tendency of dumbbells; thus, there is almost no visible effect of HI on the extension [Fig.~\ref{fig:avgR-Wi}(c)].


\section{Stretching statistics amidst HI}\label{sec:stats}

\subsection{HI steepen the coil-stretch transition}\label{sec:transition}

We
expect HI to alter the relaxation time of a chain, in a manner that depends on the number of beads. This implies that the true Weissenberg number of a chain with HI should differ from $\Wi = \lambda \tau^D$, the value for a free draining chain, by a factor that depends on $\h$ and $\Nb$. Now, if such a rescaling of the Weissenberg number was the only effect of HI, then 
the results in Fig.~\ref{fig:avgR-Wi} should show that increasing $\h$ simply translates the curves of $\Rmean$ versus $\Wi$ to the left (for dumbbells) or to the right (for chains). We now check if this is the case. We first define a critical $\Wic$ as the value of $\Wi$ at which $\Rmean$ attains half its large-$\Wi$ asymptote. We then define a rescaled $\Wistar \equiv (1/2) \Wi/\Wic$, so that the curves for all $\h$ coincide at the transition point $\Wistar = \Wistar_c = 1/2$. \revise{This rescaling is therefore an empirical shift of the Weissenberg number, based on a $\h$ dependent time scale that is determined from the results, \textit{post facto}, such that the mean extension undergoes a rapid increase at $\Wistar = 1/2$ for all values of $\h$.} \rp{One may view $\Wistar$ as the Weissenberg number that would be obtained if the $\h$ dependent time scale that controls the coil-stretch transition were used to nondimensionalize the Lyapunov exponent.} The question is whether the curves entirely collapse after this rescaling or not.

The rescaled plots of $\Rmean$ as a function of $\Wistar$ are presented in Fig.~\ref{fig:avgR-Wistar}. \rp{For dumbbells, the weak effects of HI, seen in Fig.~\ref{fig:avgR-Wi}(a), become negligible after the rescaling [see Fig.~\ref{fig:avgR-Wistar}(a)]. In contrast, the much stronger effects of HI on twenty-bead chains [Fig.~\ref{fig:avgR-Wi}(b)] cannot be accounted for by rescaling $\Wi$,} i.e, by using a HI-modified elastic relaxation time. 
Indeed, Fig.~\ref{fig:avgR-Wistar}(b) shows 
that HI steepens the coil-stretch transition, by causing chains below $\Wicstar$ to stretch less and those above $\Wicstar$ to stretch more. In other words, on comparing chains with and without HI, \rp{at the same distance from the corresponding coil-stretch transitions}, we find that HI causes stiff chains to stretch less and moderately elastic chains to stretch more.

\rp{The inset of Fig.~\ref{fig:avgR-Wistar}(c), which presents the values of $\Wi_c$ for different $\Nb$ and $\h$, shows how HI shifts the coil-stretch transition for chains with different numbers of beads. Importantly, the shift changes direction and increases in magnitude as $\Nb$ is increased from 2 to 20. Thus, while $\Wic$ decreases slightly with $\h$ for dumbbells (compare $\Wi_c$ for $\h = 0$ and 0.49, at $\Nb = 2$), it increases strongly with $\h$ for twenty bead chains ($\Nb = 20$).} (Note that we have used $\Wi_c$ to rescale $\Wi$ such that $\Wistar_c = 1/2$ at the transition because the value 1/2 marks the transition in the theory of \citet{bfl00,bfl01}. In their work, however, the transition was described in terms of the exponent of the power-law scaling of the PDF of $R$. We cannot follow the same approach here because, as we show in the next section, the PDF is qualitatively modified by HI to the extent that a clear power-law range is no longer visible. We have therefore used a criterion based on $\Rmean/\Req$, which yields a $\Wic$ that differs from 1/2 even for free-draining chains [see the inset of Fig.~\ref{fig:avgR-Wistar}(c)].)


\rp{The results in this section show that one cannot capture the effects of HI on a multibead chain by adding HI to a dumbbell. Being unable to form a physical coil, the dumbbell cannot represent the effects of hydrodynamic shielding, and so its behavior in the presence of HI departs from that of chains. The disagreement is both quantitative and qualitative: not only is the effect of HI on the extension of dumbbells much weaker than that on chains, but it also acts in the opposite direction. This contrast is evident in the variation of the location of the coil-stretch transition $\Wi_c$, which shifts slightly towards smaller $\Wi$ for dumbbells but strongly towards larger $\Wi$ for chains (inset of Fig.~\ref{fig:avgR-Wistar}(c)). Furthermore, even after accounting for this difference in the shift of the transition, by rescaling $\Wi$ to $\Wistar$, chains with HI exhibit a steepening of the transition that is absent for dumbbells [compare Fig.~\ref{fig:avgR-Wistar}(a) and Fig.~\ref{fig:avgR-Wistar}(b)].}

\rp{Hence, one has to refine the dumbbell model by adding more beads in order to obtain, even qualitatively, the behavior of chains with HI.} Quantitatively, we find, for the limited range of $\Nb$ considered here, that the effects of HI increase systematically as the number of beads are increased (i.e., as the model is refined for fixed values of the polymer's equilibrium and contour lengths, $\Rmax$ and $\Req$; see Sec.~\ref{sec:mapping}). Fig.~\ref{fig:avgR-Wistar}(d) shows that the effects of HI are much weaker in ten-bead chains when compared to twenty-bead chains [Fig.~\ref{fig:avgR-Wistar}(b)]. In fact, for a given non-zero value of $\h$, chains with more beads undergo a sharper coil-stretch transition (see Fig. S1 of the supplemental material (SM) \citep{supplement}, where the results of Fig.~\ref{fig:avgR-Wistar} are replotted to facilitate a comparison of chains with the same $\h$ but different $\Nb$). 



\begin{nolinenumbers}
\begin{figure}
\centering
\begin{subfigure}[t]{0.4\textwidth}
    \subcaption{}
    \centering
    \includegraphics{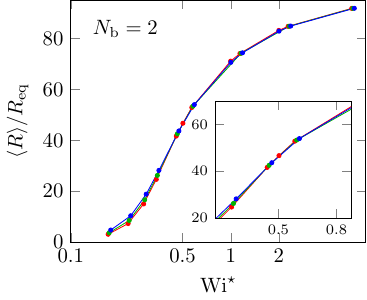}
  \end{subfigure}
  \hspace{0.2cm}
    \begin{subfigure}[t]{0.4\textwidth}
    \subcaption{}
    \centering
    \includegraphics{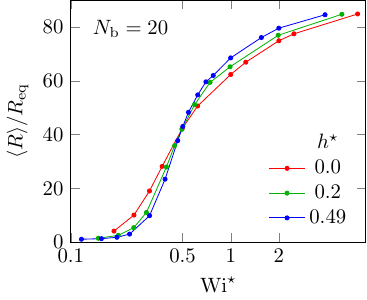}
  \end{subfigure}
  \hfill
  \begin{subfigure}[t]{0.4\textwidth}
    \subcaption{}
    \centering
    \includegraphics{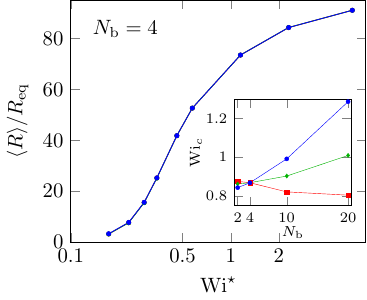}
  \end{subfigure}
  \hspace{0.2cm}
  \begin{subfigure}[t]{0.4\textwidth}
    \subcaption{}
    \centering
    \includegraphics{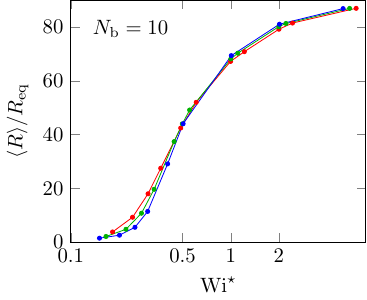}
  \end{subfigure}
\caption{\label{fig:avgR-Wistar}
Mean rescaled extension $\langle R\rangle/R_{eq}$ for various values of the rescaled Weissenberg number $\Wistar$, considering $\h$ = 0, 0.2, 0.49 [see the legend in panel (b)]: Results are presented for (a) dumbbells, (b) twenty-bead chains. (c) four-bead chains, and (d) ten-bead chains. The inset of panel (a) is a zoom of the main panel. The three curves of $\langle R\rangle/R_{eq}$ in panel (c) almost entirely overlap. The inset of panel (c) shows the critical value $\Wi_c$ of the coil-stretch transition used to rescale $\Wi$, so that $\Wicstar = 1/2$ for all $\h$. }
\end{figure}
\end{nolinenumbers}

\rp{Having examined how the coil-stretch transition is affected by HI, we proceed in the following sections to compare the statistics of extension of chains with different $\h$ and $\Nb$ at the same distance from their corresponding transition points.} To facilitate this comparison, which reveals the effects of HI that cannot be accounted for by a $\h$-dependent time-scale, we henceforth present results in terms of $\Wistar$.

\subsection{HI limit self-similar stretching}\label{sec:pdf}

Now, let us examine the distribution of polymer extension. 
Recall that, in the absence of HI, the PDF of $R$ has been shown to exhibit a power-law behavior for $\Req \lesssim R \lesssim \Rmax$, with an exponent that increases with the Weissenberg number. This result was derived theoretically for a Hookean dumbbell in a general chaotic flow \citep{bfl00,bfl01,c00}, and confirmed in simulations of FENE dumbbells and chains in turbulent flows \citep{wg10,bmpb12,rll22,sbgc22,ppv23}. We now examine whether this power-law behavior persists in the presence of HI. 

\begin{nolinenumbers}
\begin{figure}
    \centering

    \begin{subfigure}[t]{0.4\textwidth}
        \caption{}
        \centering
        \includegraphics{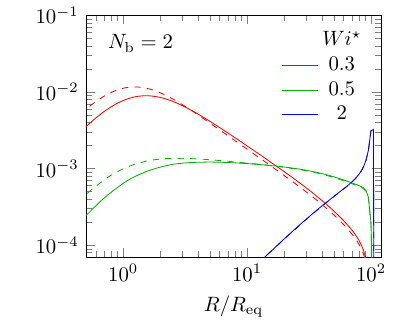}
    \end{subfigure}
    \hspace{0.5cm}
    \begin{subfigure}[t]{0.4\textwidth}
        \caption{}
        \centering
        \includegraphics{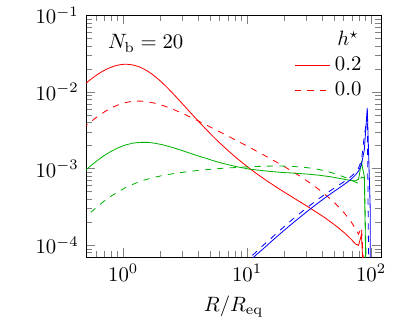}
    \end{subfigure}
    
    \caption{PDF of $\RbyReqpdf$ for $\Wistar$ = 0.3, 0.5, 2.0 [legend in panel (a)], considering (a) dumbbells, and (b) twenty-bead chains. In both plots, the solid and dashed lines correspond to cases with and without HI ($\h = 0.20$ and $\h = 0$), respectively.}
    \label{fig:pdfR-HI}
\end{figure}
\end{nolinenumbers}

The PDF of $R$ for dumbbells is presented in Fig.~\ref{fig:pdfR-HI}(a), for $\Wistar$ = 0.3, 0.5, and 2.0, both with and without HI (solid lines are for $\h=0.2$ while the dashed lines are for $\h=0$). The logarithmic axes of this plot makes apparent the power-law behaviour of the PDF for $\Req \lesssim R \lesssim \Rmax$. Importantly, we see that HI does not disrupt the power-law behaviour, though, for small and intermediate $\Wistar$, it does increase slightly the power-law exponent, so that the fraction of shrunk dumbbells ($R \sim \Req$) is marginally reduced in favour of stretched ones ($R \gg \Req$). This result is consistent with recent analytical predictions for HI-endowed dumbbells in a random chaotic flow \citep{pv25}.

\rp{The results for chains are again very different from that for dumbbells. Fig.~\ref{fig:pdfR-HI}(b), which presents the PDF of $R$ for twenty-bead chains, shows that HI produce a} 
substantial increase in the fraction of coiled polymers ($R \sim \Req$) and a corresponding reduction in the fraction of stretched polymers ($R \gg \Req$). This HI-induced modification changes the shape of the PDF, for small and intermediate values of $\Wistar$, and limits the power-law range. Taking the case of $\Wistar = 0.3$ as an example [red lines in Fig.~\ref{fig:pdfR-HI}(b)], we see that the power-law variation, which in the absence of HI held sway for $\Req \lesssim R \lesssim \Rmax$, emerges in the presence of HI only for $20\Req \lesssim R \lesssim \Rmax$. For $R > 20\Req$, the beads of the $\Nb = 20$ chain are sufficiently far apart for the strength of HI to diminish. For large $\Wistar$ almost all chains (and dumbbells) are strongly stretched and nearly unaffected by HI [see the PDF for $\Wistar = 2.0$ in Figs.~\ref{fig:pdfR-HI}(a,b)].

\begin{nolinenumbers}
\begin{figure}
    \centering

    \begin{subfigure}[t]{0.4\textwidth}
        \caption{}
        \centering
        \includegraphics{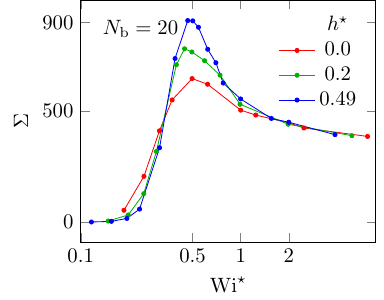}
    \end{subfigure}
    \hspace{0.2cm}
    \begin{subfigure}[t]{0.4\textwidth}
        \caption{}
        \centering
        \includegraphics{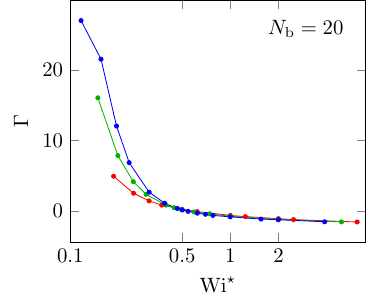}
    \end{subfigure}
    \hfill
    \begin{subfigure}[t]{0.4\textwidth}
        \caption{}
        \centering
        \includegraphics{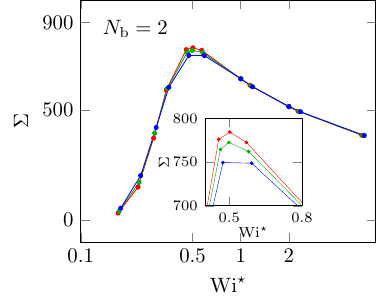}
    \end{subfigure}
    \hspace{0.2cm}
    \begin{subfigure}[t]{0.4\textwidth}
        \caption{}
        \centering
        \includegraphics{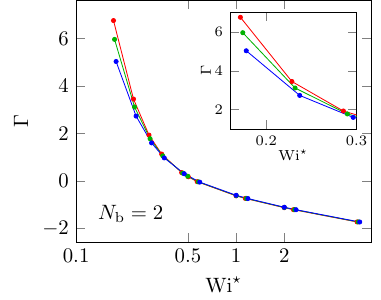}
    \end{subfigure}
  \caption{
  \revise{(a,c) Variance $\Sigma = \langle (R-\Rmean)^2 \rangle/\Req^2$  and (b,d) skewness $\Gamma = \langle (R-\Rmean)^3 \rangle/\Sigma^{(3/2)}$ of the PDF of $R$, for various values of the rescaled Weissenberg number $\Wistar$, considering $\h$ = 0, 0.2, 0.49 [see the legend in panel (a)]. Results for twenty bead chains and dumbbells are presented in the top and bottom rows, respectively. 
  The corresponding PDFs for $\Wistar = 0.3,$ 0.5, and 2.0 are shown in Fig.~\ref{fig:pdfR-HI}(b) and Fig.~\ref{fig:pdfR-HI}(a).
  }}
    \label{fig:Rvariance}
\end{figure}
\end{nolinenumbers}

The effects of HI increase in magnitude with $\h$ while remaining qualitatively the same [see Fig. S2(c) in the SM \citep{supplement} where the analogue of Fig.~\ref{fig:pdfR-HI}(b) is presented for $\h = 0.49$]. Also, as already demonstrated by the results for $\Rmean$ in Fig.~\ref{fig:avgR-Wistar}(b,d), reducing the number of beads in the chain weakens the hydrodynamic shielding effect of HI. So, the changes to the PDF of $R$ are similar but smaller for ten-bead chains, when compared to twenty-bead chains [see Fig. S2(d) in the SM \citep{supplement}]. 

\revise{Figures~\ref{fig:Rvariance}(a,b) quantify the influence of HI on the PDF of $R$, for twenty bead chains, by plotting (a) the variance $\Sigma = \langle (R-\Rmean)^2 \rangle/\Req^2$ and (b) the skewness $\Gamma = \langle (R-\Rmean)^3 \rangle/\Sigma^{(3/2)}$ as function of $\Wistar$ for different values of $\h$. In the absence of HI, the variance $\Sigma$ is known to peak near the coil-stretch transition, $\Wistar = 1/2$, where the distribution is at its broadest~\citep{mav05,gcs05,wg10}. We see the same behavior of $\Sigma$ in the presence of HI; the peak at $\Wistar = 1/2$ becomes more prominent as $\h$ increases because the corresponding PDF develops a bimodal shape, with local peaks near $\Req$ and $\Rmax$ [see the dashed and solid green lines in Fig.~\ref{fig:pdfR-HI}(b), as well as the blue lines in Figs.~\ref{fig:Nbvary-pdf}(b) and~\ref{fig:Nbvary-pdf}(d) below]. The restriction by HI of the power-law range of the PDF of $R$ for stiff chains, noted earlier, is associated with an increase in the skewness of the distribution [see the red curves for $\Wistar = 0.3$ in Fig.~\ref{fig:pdfR-HI}(b)]; Fig.~\ref{fig:Rvariance}(b) quantifies the extent of this effect showing a strong increase in $\Gamma$ with $\h$ for small values of $\Wistar$.} 

\rp{In the case of dumbbells [Figs.~\ref{fig:Rvariance}(c,d)], $\Sigma$ and $\Gamma$ exhibit a much weaker dependence on HI as compared to twenty bead chains; moreover, the changes occur in the opposite direction, so that $\Sigma$ and $\Gamma$ decrease slightly with $\h$ in the case of dumbbells [see the insets of Figs.~\ref{fig:Rvariance}(c,d)], whereas they increase strongly with $\h$ in the case of twenty-bead chains [Figs.~\ref{fig:Rvariance}(a,b)]. Thus, adding HI to a dumbbell fails to capture the effects of HI on the extension of multibead chains, with regard to not only the mean [Fig.~\ref{fig:avgR-Wistar}(a,b)] but also the distribution. }

\begin{nolinenumbers}
\begin{figure}
    \centering
    \begin{subfigure}[t]{0.37\textwidth}
        \caption{}
        \centering
    \includegraphics{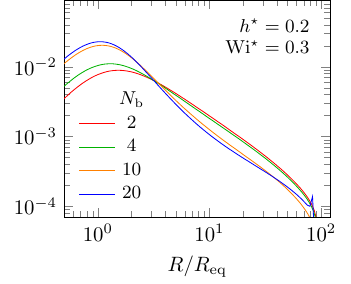}
    \end{subfigure}
    \hspace{0.5cm}
    \begin{subfigure}[t]{0.37\textwidth}
        \caption{}
        \centering
    \includegraphics{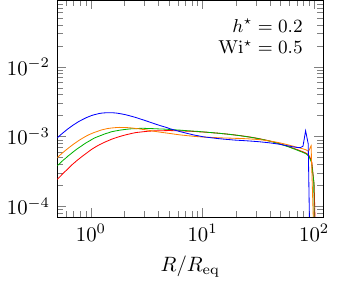}
    \end{subfigure}
\hfill
    \begin{subfigure}[t]{0.37\textwidth}
        \caption{}
        \centering
    \includegraphics{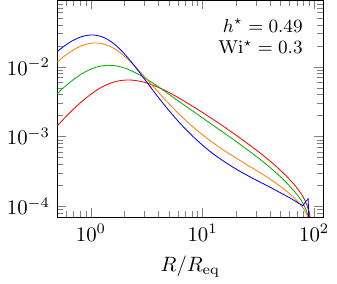}
    \end{subfigure}
    \hspace{0.5cm}
    \begin{subfigure}[t]{0.37\textwidth}
        \caption{}
        \centering
    \includegraphics{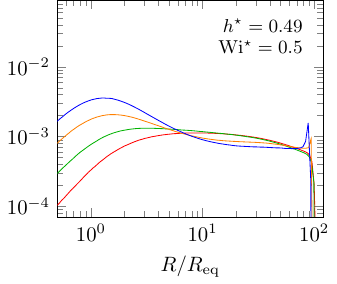}
    \end{subfigure}
\caption{PDF of $R/\Req$ for chains with different numbers of beads $\Nb$, and $\h = 0.2$ (top row) or $\h = 0.49$ (bottom row). The results are presented for $\Wistar = 0.3$ and $0.5$.}
\label{fig:Nbvary-pdf}
\end{figure}
\end{nolinenumbers}

Figure~\ref{fig:Nbvary-pdf} depicts how the PDF of $R$ changes on increasing the number of beads, while keeping $\h$ fixed. 
\rp{The PDF changes systematically with $\Nb$, such that the power-law distribution for dumbbells ($\Nb = 2$) is altered substantially once $\Nb$ reaches 20. At small $\Wistar$, increasing $\Nb$ produces an increase in the fraction of small-extension configurations [Figs.~\ref{fig:Nbvary-pdf}(a,c)] due to the action of hydrodynamic shielding which grows in relevance with $\Nb$. As a consequence, the power-law behavior is progressively disrupted and the skewness of the PDF increases; indeed, $\Gamma$ in Figs.~\ref{fig:Rvariance}(b,d) is much larger for $\Nb = 20$ than for $\Nb = 2$, at small $\Wistar$. Near the coil-stretch transition, the PDF develops a bimodal character as $\Nb$ is increased [Figs.~\ref{fig:Nbvary-pdf}(b,d)], which in turn increases the variance; thus, $\Sigma$ in Figs.~\ref{fig:Rvariance}(a,c) is much larger for $\Nb = 20$ than for $\Nb = 2$, for $\Wistar \approx 1/2$. The extent of these changes with $\Nb$ is greater for larger values of $\h$ [compare Figs.~\ref{fig:Nbvary-pdf}(a,b) with Figs.~\ref{fig:Nbvary-pdf}(c,d)].} 

It is natural to ask how much further the stretching behavior of chains would change if one were to continue increasing $\Nb$ up to the number of Kuhn steps $\Nk$. This question---about the asymptotic dependence on the number of beads of the dynamics of a chain with HI---is key to the development of successive fine graining schemes for accurately modeling polymers \citep{pps04}; it has been well-studied in the context of equilibrium properties \cite{jendrejack02} and viscometric flows \citep{Hsieh03,Sunthar05,pps04,psp04,praphul24}. To answer this question in turbulent flows would require increasing $\Nb$ well beyond twenty, a computationally demanding and important task for future work. \revise{In the absence of a fine-graining scheme for chains with HI in turbulence, we do not know how to adjust $\h$, as $\Nb$ is increased, so that the different chains correspond to increasingly refined models of the same polymer. We have therefore kept $\h$ fixed and varied $\Nb$ in Fig.~\ref{fig:Nbvary-pdf} to show
how the sensitivity of the extensional dynamics to HI increases with the number of beads.}

\subsection{HI delay migrations between coiled and stretched states}\label{sec:persistence}

A characteristic feature of HI-endowed chains in extensional flow is the appearance of a coil-stretch hysteresis \citep{sbsc03,ssc04}: the transition from coiled to stretched configurations when $\Wi$ is increased occurs at a larger value of $\Wi$ than that at which the reverse transition occurs when $\Wi$ is decreased. For $\Wi$ values within the hysteresis window, the effective conformation-dependent drag introduced by HI causes chains to persist in either stretched or coiled states, i.e., HI significantly increase the typical time required for Brownian fluctuations to induce a transition between these states \citep{sbsc03,cpv06}. Therefore, when the mean extension is measured after a finite time of observation, its value will depend on whether the polymers were initially coiled or stretched. 

\begin{nolinenumbers}
\begin{figure}
    \centering
        \includegraphics[width = \textwidth,height = 0.25\textwidth,keepaspectratio]{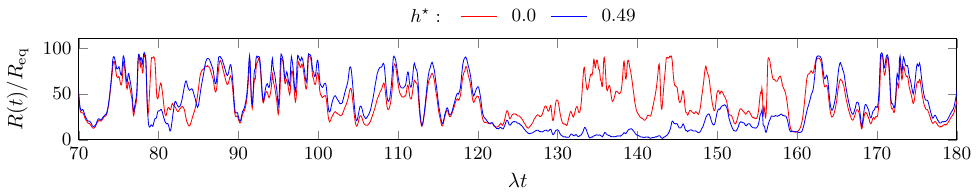}
\caption{
Evolution of $R(t)$ along a typical trajectory in the turbulent flow, within the stationary regime, for twenty bead chains with and without HI ($\h = 0.49$ and 0.0) and for $\Wistar = 0.5$.}
\label{fig:Rtrace}
\vspace{\baselineskip}
    \centering
    \begin{subfigure}[t]{0.3\textwidth}
        \caption{}
        \centering
        \includegraphics{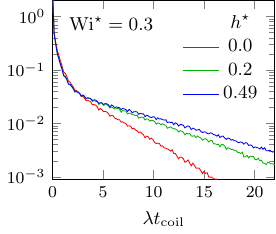}
    \end{subfigure}
    \begin{subfigure}[t]{0.3\textwidth}
        \caption{}
        \centering
        \includegraphics{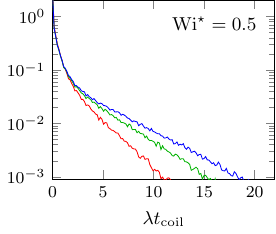}
    \end{subfigure}
    \begin{subfigure}[t]{0.3\textwidth}
        \caption{}
        \centering
        \includegraphics{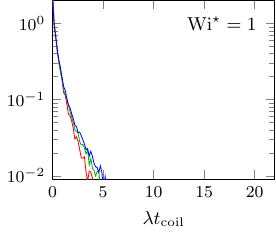}
    \end{subfigure}

    \begin{subfigure}[t]{0.3\textwidth}
        \caption{}
        \centering
        \includegraphics{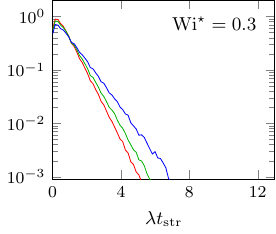}
    \end{subfigure}
    \begin{subfigure}[t]{0.3\textwidth}
        \caption{}
        \centering
        \includegraphics{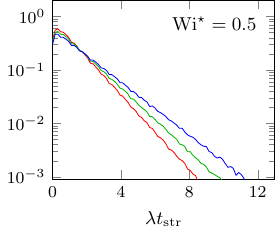}
    \end{subfigure}
    \begin{subfigure}[t]{0.3\textwidth}
        \caption{}
        \centering
        \includegraphics{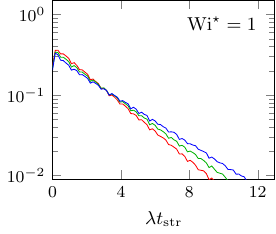}
    \end{subfigure}
\caption{
PDF of the persistence time of twenty-bead chains in (a-c) coiled states ($R >5 \Req$) and in (d-f) stretched states ($R > 50 \Req$). Results are presented for $\Wistar = 0.3$, 0.5, and 1.0, and for $\h$ = 0, 0.2, 0.49 [legend in panel(a)].}
\label{fig:persist}
\end{figure}
\end{nolinenumbers}

Now in turbulence, the strong fluctuations of the velocity gradient cause polymers to sample a wide range of extensions and to migrate between states of large and small extension relatively quickly.
The evolution of the end-to-end extension is illustrated in Fig.~\ref{fig:Rtrace}, which presents a typical time-trace of $R(t)$ for a twenty-bead chain ($\Wistar = 0.5$), with and without HI (the same Lagrangian trajectory, and hence the same time series for $\bm \kappa(t)$, is used for both cases).
We see that, even with HI, chains repeatedly stretch and recoil as they move through the turbulent flow. Therefore, after an initial transient (which lasts less than $\sim40\lambda t$ for the range of $\Wistar$ considered here), a stationary regime is attained whose statistics are independent of the initial configuration. Hence, HI cannot induce a coil-stretch hysteresis in the extension statistics measured in turbulence \citep{cpv06}. However, it is possible that HI modify the typical time it takes for chains to migrate between states of small and large extension. Indeed, Fig.~\ref{fig:Rtrace} shows that the chain with HI attains both smaller and larger extensions than the chain without HI. The HI-endowed chain experiences a prolonged episode with very small extensions. However, when both chains are stretched by the flow, the fluctuations towards smaller extensions seem to be weaker for the HI-endowed chain. 

To quantitatively investigate the migration between small and large extensions, we define two thresholds, $\ell_\mathrm{coil}$ and $\ell_\mathrm{str}$, and designate chains with $R < \ell_\mathrm{coil}$ as coiled and those with $R > \ell_\mathrm{str}$ as stretched. We then compute the intervals of time for which chains remain in these two states, and thereby obtain the PDF of persistence times in both states. We use $\ell_\mathrm{coil} = 5\Req$ and $\ell_\mathrm{str} = 50 \Req$, after checking that our conclusions do not depend on the precise values of these thresholds [see the SM~\citep{supplement}]. A similar analysis of the persistence time in stretched states was performed for Rouse chains (no HI) in \citet{ppv23}, where the corresponding PDF was found to have an exponential tail. 

Figures~\ref{fig:persist}(a-c) present the PDFs of the persistence time in coiled states, $t_\mathrm{coil}$, for twenty-bead chains with (a) $\Wistar = 0.3$, (b) $\Wistar = 0.5$, and (c) $\Wistar = 1.0$. \revise{In the presence of HI, the PDF continues to have an exponential tail and broadens so that the probability of experiencing larger values of $t_\mathrm{coil}$ increases significantly as $\h$ is increased.} The same is true for the PDFs of the persistence time in stretched states, $t_\mathrm{str}$, as seen in Figs.~\ref{fig:persist}(d-f). Clearly, HI does increase the time for migrating between coiled and stretched states. This effect is inline with HI inducing an effective conformation-dependent drag. 


In Fig.~S3 of the SM, we present the persistence time PDFs for ten-bead chains and for dumbbells \citep{supplement}. As expected from the discussion in Sec.~\ref{sec:db-vs-chain}, the results for dumbbells are opposite to those for chains, with HI-endowed dumbbells taking less time to migrate between shrunk and stretched states. The results for ten-bead chains are similar to those for twenty-bead chains, but with weaker effects of HI owing to the reduction of the number of beads.

\section{Random gradients emulate turbulence}\label{sec:random}

Before concluding, we check whether the effects of HI on turbulent polymer stretching can be predicted by using random Gaussian velocity gradients, i.e., by replacing the turbulent flow with a synthetic random flow. As discussed in Sec.~\ref{sec:turbflow}, the time series of random gradients is constructed so as to have the same Lyapunov exponent and similar temporal correlation properties as the Lagrangian trajectories from the DNS (see Appendix \ref{sec:turb-details} for details). If the Gaussian gradients work well, then one can study the extensional dynamics of polymer chains without having to perform a DNS. This would facilitate the development of coarse-grained models, like dumbbells with a conformation-dependent drag \citep{deGennes74,Hinch77,prabhakar-blob-07,cpv06}, by allowing for the comparison of polymer models in a fluctuating flow without having to simulate the Navier-Stokes equations. Data-driven modelling approaches \citep{Sindy,Sapsis2018,Danny,Graham24-data-fibre} which require large amounts of data would particularly benefit from using the random flow model, since one can then perform very long simulations without the need for long Lagrangian trajectories from a DNS.

The random Gaussian gradient model has already been shown to work well for Rouse chains (without HI) by \citet{ppv23}. We now check whether this remains the case for chains with HI. We focus on twenty-bead chains, for which HI produce prominent changes in the distribution of extension (Fig.~\ref{fig:pdfR-HI}).

The PDF of extension predicted using the Gaussian gradients (GG) are compared to the results from the DNS, for twenty-bead chains without HI in Fig.~\ref{fig:DNS-GG-pdf}(a). The comparison with HI is presented in Fig.~\ref{fig:DNS-GG-pdf}(b) for $\h = 0.2$ and in Fig.~\ref{fig:DNS-GG-pdf}(c) for $\h = 0.49$. The DNS and GG results agree reasonably well even in the presence of HI. Importantly, the differences between the PDFs from the DNS and the GG model are much smaller than the modifications produced by HI [compare the differences in Fig.~\ref{fig:DNS-GG-pdf}(b) to those in Fig.~\ref{fig:pdfR-HI}(b)]. This means that we can replace the DNS by the GG random model without compromising our ability to study and understand the effects of HI on polymer stretching in fluctuating flows. 

After averaging over the PDFs, the results for the mean extension $\Rmean/\Req$ from the DNS and the GG random model match well, as shown in Fig.~S4 of the SM \citep{supplement}. We have also found that the effects of HI on the persistence time in stretched and coiled states (discussed in Sec.~\ref{sec:persistence}) are reproduced by the GG random model.

\revise{It is important to note that, though the principal Lyapunov exponent $\lambda$ is the same in the two flows, the distribution of $\nabla \bm{u}$ is very different. The velocity-gradient in turbulence is well-known to exhibit strong extreme-valued fluctuations~\citep{Buaria2019,Buaria2022} that will be absent in the Gaussian model. Indeed, we find that the PDF of the strain-rate in the turbulent DNS has strongly non-Gaussian, flared tails and differs markedly from the corresponding PDF in the Gaussian flow (see Fig.~S8 in the SM~\citep{supplement}). The close agreement of the extension statistics in these two flows implies, therefore, that the distribution of $\nabla \bm{u}$, particularly, the presence of extreme events, does not strongly affect the statistics of polymer stretching.
As explained in Ref.~\citep{ppv23}, this lack of sensitivity to extreme straining is a consequence of the elasticity of the polymer: the polymer extension does not depend on the instantaneous strain-rate but rather effectively integrates the fluctuating strain rate over a time period that scales with the elastic relaxation time.}

\revise{The relative unimportance of extreme strain-rate events 
has implications for how individual polymer stretching is altered by elastic feedback forces through its modification of the global flow. In this work, we have accounted for the local effect of feedback forces by including HI but neglected the global effect, which will change the distribution of the fluctuating velocity gradient experienced by the polymer.
In homogeneous isotropic turbulence, elastic feedback reduces both the mean magnitude of the strain-rate and its extreme fluctuations~\citep{bfl00,bfl01,prasad10,rll22}. The results of this section suggest that the loss of extreme events will not matter and that polymer extension will be affected primarily by $\lambda$. This change can be accounted for by adjusting $\Wi$, and so we expect that the distribution of polymer extension in the presence of global elastic feedback will correspond closely to that for a lower $\Wi$ without global feedback. This has been demonstrated in two-way coupled simulations of dumbbells without HI by Ref.~\citep{rll22}. Our present results suggest that the same behavior will be obtained for chains with HI.}

\begin{nolinenumbers}
\begin{figure}
\centering
\begin{subfigure}[t]{0.325\textwidth}
    \captionsetup{margin={-5pt,0pt}}  
    \caption{}
\centering
    \includegraphics{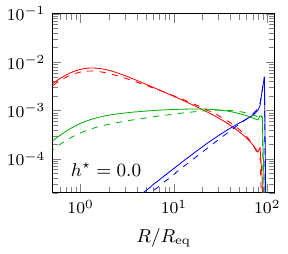}
\end{subfigure}
\begin{subfigure}[t]{0.325\textwidth}
    \captionsetup{margin={-5pt,0pt}}  
    \caption{}
\centering
    \includegraphics{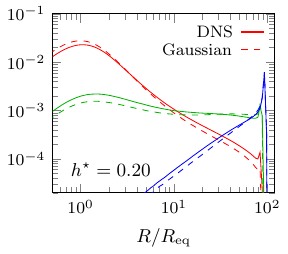}
\end{subfigure}
\begin{subfigure}[t]{0.325\textwidth}
    \captionsetup{margin={-5pt,0pt}}  
    \caption{}
\centering
    \includegraphics{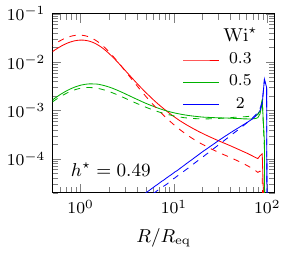}
\end{subfigure}
\caption{PDF of $\RbyReqpdf$ for twenty bead chains stretched by velocity gradients from the DNS of turbulence (solid line) and from the Gaussian-gradient random model (dashed line). The comparison is presented for (a) $\h = 0$ (no HI), (b) $\h = 0.2$, and (c) $\h = 0.49$.}
\label{fig:DNS-GG-pdf}
\end{figure}
\end{nolinenumbers}


\section{Concluding remarks}\label{sec:conclusion}

Simulations of viscoelastic turbulence typically adopt dumbbell-based descriptions of the
polymer, either indirectly through continuum constitutive equations such as Oldroyd--B or FENE-P~\citep{Hinch21}, or
directly through Eulerian--Lagrangian (EL) approaches in which explicit dumbbells are evolved alongside
the Navier--Stokes equations~\citep{Keunings04,wg13,sbgc22,rll22,sbgc2025-dist}. \rp{A central modeling
question is what is lost in such very coarse, fixed-drag descriptions when the underlying polymer is a
long flexible chain for which intramolecular hydrodynamic interactions (HI) generate shielding and hence
conformation-dependent hydrodynamic coupling.}
To gain insight into this issue---raised decades ago in the context of turbulent flows by Hinch~\citep{Hinch77}---we have 
examined the effect of HI on the stretching of a polymer in turbulence, considering a range of $\Wi$ and different levels of coarse-graining (from dumbbells to twenty bead chains).

\rp{Within the dumbbell idealization ($\Nb=2$), we find that adding explicit HI produces negligibly small changes in the stretching statistics. However, this \emph{should not} be interpreted as validating fixed-drag
dumbbell models as faithful surrogates for physically realistic chains with intramolecular HI. Instead,
our fine-graining results show that once the polymer is allowed to form physically coiled configurations,
HI induces hydrodynamic shielding and thereby slows migration between coiled and stretched states. This
mechanism reorganizes the stretching dynamics in a way that cannot be reduced to a simple horizontal shift
of the $\Wi$ axis: it steepens the coil--stretch transition and reshapes the PDF of extension at small and
intermediate $R$. For a fixed nonzero value of the HI parameter $h^\star$, the deviation between HI-endowed and free-draining predictions
 is most pronounced near
$\Wistar\simeq 1/2$, where polymers repeatedly traverse coiled and stretched configurations. At large
$\Wistar$, where chains spend most of their time strongly stretched and inter-bead separations are large,
the influence of HI on mean extension becomes comparatively weak. }

These effects of HI grow systematically with $\Nb$, over the modest range of two to twenty considered in this study.
We expect the influence of HI to continue to increase with further fine-graining, i.e, as $\Nb$ is increased toward the Kuhn-step resolution $\Nk$, until a saturation is attained for a large number of beads. \rp{The
asymptotic $\Nb$-dependence of HI-endowed chain dynamics is central to successive fine-graining schemes
\citep{pps04}, which are well developed for equilibrium properties \citep{jendrejack02} and viscometric
flows \citep{Hsieh03,Sunthar05,pps04,psp04,praphul24}. Extending such ideas to turbulence would require
simulations at $\Nb\gg 20$ or the development of extrapolation strategies based on how leading-order
corrections scale with $\Nb$ at fixed $\Nk$.} Understanding this scaling in turbulent flows is an important
direction for future work.

The most obvious characteristic of multibead chains, which is absent in dumbbells, is that elastic relaxation occurs on multiple time-scales. However, this feature alone does \textit{not} lead to differences in the dynamics of the end-to-end extension. Indeed, the transient and stationary statistics of a Rouse chain in a homogeneous isotropic turbulent flow can be mapped to that of a dumbbell using the Jin-Collins mapping \citep{wg10,ppv23}. Our current work has shown that this simple equivalence does not hold once HI are introduced. While stiff dumbbells stretch slightly more due to HI, stiff chains stretch less (the contrast reverses at higher elasticity). Furthermore, while the distribution of extension for dumbbells continues to exhibit a power-law, from its equilibrium value to its maximum value, the shape of the distribution for chains is strongly modified such that a power-law range is not clearly identifiable. \rp{The important implication of these results is that one cannot capture, not even qualitatively, the effects of HI on a polymer by simply adding HI to a dumbbell.}

\rp{Now, the elastic dumbbell certainly is and will remain a valuable model; it enables computationally tractable simulations of viscoelastic turbulence by providing a minimal representation of the essential physics of flow-induced stretching and entropic relaxation.
However, if the goal is predictive capability for engineering design---for example, to predict
the extent of drag reduction achievable by a given polymer of a certain molecular weight---then our results show that the dumbbell is inadequate because it misses systematic chain-level effects of HI.}
This provides motivation for
developing reduced-order models that remain computationally tractable yet emulate chain-level
hydrodynamics, for example dumbbells endowed with an extension-dependent (or otherwise conformation-aware)
drag \citep{deGennes74,Hinch77,cpv06,prabhakar-blob-07}. To calibrate such models, it would be helpful to perform large-ensembles of simulations to generate fitting data or to compare polymer models; the Gaussian random flow will be helpful in this regard by serving as a surrogate for the turbulent flow.

In the context of dumbbells vis-\`a-vis chains, the EL simulations of \citet{sbgc2024-chains} show that free-draining (no HI) large-$\Wi$ FENE chains get trapped into folded states in \textit{non-isotropic}, shear-dominated turbulent pipe flow. Consequently, chains exhibit multiple spurious peaks in the distribution of the end-to-end extension. The only coarse-grained model that avoids this artifact in turbulent pipe flow is the dumbbell, simply because of its inability to fold. Now, our present results suggest that if HI were to be included then the dumbbell would not yield qualitatively correct predictions, precisely because it cannot fold and form a physical coil (see Sec.~\ref{sec:db-vs-chain}). The resolution to this predicament could be to use a modified dumbbell with extension-dependent drag, as discussed above. This is an interesting direction for future work, with the natural next step being an analysis of the dynamics of HI-endowed chains in turbulent shear flows. [Note: In EL simulations, a feedback force is applied by each bead onto the fluid, but the typical spatial resolution of the flow solver (chosen to resolve the smallest scales of the turbulent flow) is insufficient to resolve the inter-bead disturbance-flow that would give rise to HI. Hence, unless HI is explicitly included in the evolution equation of the dumbbell/chain, HI effects will not be captured by the EL simulation.] 

In homogeneous isotropic turbulence that is devoid of mean shear, chains with and without HI \citep{wg10,vwrp21,ppv23} do not exhibit persistent folded configurations, as demonstrated by their smooth distributions of extension. 
The strong influence on the behaviour of chains of the anisotropy of the turbulent flow offers an interesting counter-point to the mild influence of the non-Gaussian distribution of strain-rates (evidenced by the success of the Gaussian random flow in emulating isotropic turbulence with and without HI \citep{ppv23}). It appears that some statistical features of the flow---degree of isotropy---do strongly affect chain stretching, while other features---non-Gaussian extreme straining---do not. With this in mind, it would be interesting to study the dynamics of chains in an anisotropic Gaussian random flow, and to examine how the likelihood of persistent folded configurations depends on the degree of anisotropy.
More generally, the possible effect of the statistical properties of the flow on the stretching of polymers should be borne in mind while developing multiscale models for turbulent polymer solutions, and models should therefore be tested in different canonical turbulent flows.

\acknowledgments
We thank Samriddhi Sankar Ray (ICTS, Bengaluru) for sharing his database of Lagrangian trajectories in homogeneous and isotropic turbulence. A.G. gratefully acknowledges the PhD Fellowship received from the IITB-Monash Research Academy (year 1) and from the Prime Minister's Research Fellowship Scheme, Ministry of Education, Govt. of India (years 2-4). The work was supported by the Indo–French Centre for the Promotion of Advanced Scientific Research (IFCPAR/CEFIPRA, project no. 6704-1). J.R.P. acknowledges his Associateship with the International Centre for Theoretical Sciences (ICTS), Tata Institute of Fundamental Research, Bengaluru,
India. D.V. acknowledges the support of Agence Nationale de la Recherche through Project
No.~ANR-21-CE30-0040-01.
Simulations were performed on the IIT Bombay workstations \textit{Gandalf} (procured through DST-SERB grant SRG/2021/001185) and \textit{Aragorn} (procured through the IIT-B grant RD/0519-IRCCSH0-021).

\appendix

\section{DNS data and the random model for velocity gradients}\label{sec:turb-details}

The DNS corresponds to stationary homogeneous isotropic turbulence, with a Kolmogorov time-scale $\tau_\eta = 3.72\times 10^{-2}$ (in simulation units). Note that $\tau_\eta$ is the turnover time-scale of the small dissipative eddies in the flow. Relative to this timescale, the Lyapunov exponent of the tracer trajectories (used for defining $\Wi$ in Sec.~\ref{sec:BD}) is found to be \revise{$\lambda=(0.136\pm 0.0006)/\tau_\eta$}. We have computed the Lyapunov exponent from the velocity gradient data using the continuous $QR$ method \citep{gpl90,drv97,mcdh01}, \revise{of which a brief description follows. 
Consider
a line element $\bm\ell(t)$ evolving along a fluid trajectory $\bm x(t)$. The Lyapunov exponent is defined as $\lambda = 
\lim_{t\to\infty} t^{-1} \vert\ell(t)/\ell(0)\vert$.
In a chaotic flow, $\lambda>0$ and therefore $\ell(t)$ grows as $\ell(t)\sim\ell(0)\mathrm{e}^{\lambda t}$ as $t\to\infty$. Thus, $\lambda$ measures the asymptotic growth rate of line elements \cite{bjpv98}.
In principle, $\lambda$ could be calculated directly by solving the evolution equation for $\bm\ell(t)$:
\begin{equation}
\label{eq:ell}
\frac{\mathrm{d}\bm \ell}{\mathrm{d}t} = \mathbb{A}(t)\bm\ell
\end{equation}
with $\mathbb{A}_{ij}(t)=\partial_j u_i(\bm x(t),t)$. However, this approach is impractical in a numerical calculation, precisely because $\ell(t)$ grows exponentially. The $QR$ method is one of the methods that have been designed to circumvent this issue.
Since \eqref{eq:ell} is a system of linear ordinary differential equations, its 
solution has the form $\bm\ell(t)=\mathbb{W}(t)\bm\ell(0)$, where $\mathbb{W}(t)$ is the fundamental matrix of the system. Therefore, the Lyapunov exponent can also be written as $\lambda = 
\lim_{t\to\infty} t^{-1} \vert\mathbb{W}(t)\hat{\bm\ell}(0)\vert$. In the $QR$ method, $\mathbb{W}(t)$ is decomposed into an orthonormal matrix $\mathbb{Q}(t)$ and an upper triangular matrix $\mathbb{R}(t)$, with positive diagonal entries. While the entries of $\mathbb{R}(t)$ can grow in time, $\mathbb{Q}(t)$ is necessarily bounded. The continuous $QR$ algorithm solves an ordinary differential equation for $\mathbb{Q}$ along the trajectory $\bm x(t)$ and recovers $\lambda$ from the temporal evolution of $\mathbb{Q}(t)$ and $\mathbb{A}(t)$ alone; any manipulation of exponentially growing quantities is thus avoided (see Refs. \citep{gpl90,drv97,mcdh01} for further details). 
}

We have also calculated the Lagrangian correlation time-scales of the strain-rate and vorticity, $\tau_{S}$ and $\tau_\varOmega$, which will be needed to construct the random velocity gradient time series (as discussed below).  Defining the rate-of-strain and rotation tensors as
$\bm{S}=(\nabla\bm u+\nabla\bm u^\top)/2$ and
$\bm{\varOmega}=(\nabla\bm u-\nabla\bm u^\top)/2$, we calculate the
autocorrelation functions of $S_{11}$ and $\varOmega_{12}$ and find an approximately exponential
decay in time. On integrating these functions, we obtain 
$\tau_{S}=2.20\,\tau_\eta$ and $\tau_\varOmega=8.89\,\tau_\eta$,
in agreement with previous numerical simulations at comparable $R_\lambda$ \citep{y01}. The value of the Kolmogorov time-scale $\tau_\eta$, mentioned above, is determined from $S_{11}$, using isotropy, as
$\tau_\eta=\left({15\langle S_{11}^2\rangle}\right)^{-1/2}$.


The random model for the velocity gradient $\bm \kappa(t)$ is adopted from \citet*{bkl97}. We have $\bm\kappa(t)=\bm{S}(t)+\bm{\varOmega}(t)$ with
\begin{equation}
\bm{S}=\sqrt{3}\,A
\begin{pmatrix}
\frac{2\zeta_1}{\sqrt{3}} & \zeta_3 & \zeta_4
\\
\zeta_3 & -\frac{\zeta_1}{\sqrt{3}}+\zeta_2 & \zeta_5
\\
\zeta_4 & \zeta_5  & -\frac{\zeta_1}{\sqrt{3}}-\zeta_2
\end{pmatrix},
\qquad
\bm{\varOmega}=\sqrt{5}\,A
\begin{pmatrix}
0 & \varpi_1 & \varpi_2
\\
-\varpi_1 & 0 & \varpi_3
\\
-\varpi_2 & -\varpi_3  & 0
\end{pmatrix},
\end{equation}
where $\zeta_i(t)$ ($i=1,\dots,5$)
and $\varpi_i(t)$ ($i=1,2,3$)
are independent zero-mean unit-variance
Gaussian random variables, with exponentially decaying autocorrelation 
functions and integral times $\taus$ and $\tau_\varOmega$, respectively. Consequently, $S_{ij}$ and $\varOmega_{ij}$ are Gaussian matrices that satisfy
$\langle S_{ij}\rangle=\langle\varOmega_{ij}\rangle=0$,
\begin{align}
\langle S_{ik}(t)S_{jl}(0)\rangle &= 3A^2
\left(\delta_{ij}\delta_{kl}+\delta_{il}\delta_{jk}-\frac{2}{3}\delta_{ik}
\delta_{jl}\right) \, e^{-t/\taus}, \\
\langle \varOmega_{ik}(t)\varOmega_{jl}(0)\rangle &= 
5A^2 \left(\delta_{ij}\delta_{kl}-\delta_{il}\delta_{jk}\right)
e^{-t/\tau_\varOmega},
\end{align}
and $\langle\varOmega_{ij}\varOmega_{ij}\rangle=\langle S_{ij}S_{ij}\rangle$. (The latter reproduces the well-known relation between the vorticity $\omega$ and the energy dissipation rate $\epsilon$: $\nu\langle\omega^2\rangle=\langle\epsilon\rangle$ \citep{f95}.)

By setting the integral times $\taus$ and $\tau_\varOmega$ to match the corresponding values of the Lagrangian DNS database, we obtain a fluctuating time series $\bm \kappa(t)$ whose temporal correlation approximates that of the turbulent velocity gradient (the correlated random numbers $\zeta_i(t)$
and $\varpi_i(t)$ are generated using the algorithm of \citet{Fox88}). 
The magnitude of the random velocity gradient is set by the constant $A$, which therefore also determines the Lyapunov exponent $\lambda$. Choosing $A= 2.538$ yields approximately the same value for $\lambda$ as in the DNS.

\FloatBarrier
\bibliography{polymers}

\end{document}